\documentclass[reprint, twocolumn, twoside,  nofootinbib, hidelinks, colorlinks, linkcolor=blue, citecolor=blue, urlcolor=black,
 amsmath,amssymb,
pre,
floatfix,
longbibliography,
]{revtex4-1}

\usepackage[switch,columnwise]{lineno}
\usepackage{fancyhdr}
\usepackage{helvet}
\usepackage{graphicx}
\usepackage{dcolumn}
\usepackage{bm}
\usepackage{hyperref}
\usepackage{mathrsfs}
\usepackage{amsbsy}
\usepackage{amsmath}
\usepackage{scrextend}
\usepackage[export]{adjustbox}
\usepackage{lipsum}

\begin{document}

\preprint{APS/123-QED}

\title{Closed, Two Dimensional Surface Dynamics}

\author{David V. Svintradze}
 \email{Present address: Max Plank Institute for the Physics of Complex System, 01187 Dresden, Germany. \\ dsvintra@yahoo.com, dvs@pks.mpg.de}
\affiliation{%
 Department of Physics, Tbilisi State University, Chavchavadze Ave. 03, 0179 Tbilisi, Georgia
}%

\date{\today}

\begin{abstract}
We present dynamic equations for two dimensional closed surfaces and analytically solve it for some simplified cases. We derive final equations for surface normal motions by two different ways. The solution of the equations of motions in normal direction indicates that any closed, two dimensional, homogeneous surface with time invariable surface energy density adopts constant mean curvature shape when it comes in equilibrium with environment. As an example, we apply the formalism to analyze equilibrium shapes of micelles and explain why they adopt spherical, lamellar and cylindrical shapes. We show that theoretical calculation for micellar optimal radius is in good agreement with all atom simulations and experiments. 
\end{abstract}

\maketitle

\section{Introduction}


Biological systems exhibit a variety of morphologies and experience large shape deformations during a motion. Such 'choreography' of shape motility is characteristic not only for all living organisms and cells \cite{doi:10.1146/annurev.immunol.24.021605.090620} but also for proteins, nucleic acids and to all biomacromolecules in general. Shape motility, which is a motion of two-dimensional surfaces, may be a result of active (by consuming energy) or passive (without consuming energy) processes. The time scale for shape dynamics may vary from slow (nanometer per nanoseconds) to very fast (nanometer per femtosecond) \cite{doi:10.1021/jp0213506, 10.2307/2666245}.  Slowly moving surfaces are considered as over-damped systems. An example is cell motility. In that case one may use well developed the Helfrich formalism to describe the motion. This is a coarse-grained description of membranes with an expansion of the free energy in powers of the curvature tensor \cite{helfrich1973elastic}. However, while the formalism  \cite{helfrich1973elastic} are applicable to slowly moving surfaces 
they are not applicable to fast moving surfaces, where biomolecules maybe fitted. Surface dynamics for proteins or DNA \cite{doi:10.1021/jp0213506, 10.2307/2666245} may reach $nm/fs$ range.
 So that surfaces may be represented as virtual three dimensional pseudo Riemannian manifolds. We derived fully generic equations of motions for three manifolds 
  \cite{2016arXiv160907765S}, but purposefully omitted lengthy discussion about motion of two-dimensional surfaces, which is a topic for this paper. 

Currently, significant progress on fluidic models of membrane dynamics has already been made. The role of geometric constraints in self-assembly have been elucidated by linking together thermodynamics, interaction free energies and geometry   \cite{Israelachvili1976, Israelachvili1977}. The Helfrich formalism provides the foundation for a purely differential geometric approach whereby the  membrane surface potential energy density is considered as a functional of the static curvature \cite{helfrich1973elastic}, see also review papers\cite{deserno2015fluid, lipowsky1995, seifert1997configurations}. The model has been improved by adding force and torque balance equations \cite{0305-4470-35-30-302, B701952A}. Specific dynamical equations accounting for bending as well as electrodynamic effects have also been reported \cite{SCRIVEN196098, PhysRevE.79.031915, GAO20082844}. Furthermore, active membrane
theories have extended our understanding of passive membranes. Active membrane theories include external forces \cite{PhysRevE.96.032404, PhysRevLett.84.3494, PhysRevLett.92.168101, PhysRevLett.93.268104} and provide a framework for the study of active biological or chemical processes at surfaces, such as the cell cortex, the mechanics of epithelial tissues, or reconstituted active systems on surfaces \cite{PhysRevE.96.032404}.

Among the remarkable aspects of fluid lipid membranes deduced from the large body of theoretical work \cite{deserno2015fluid, lipowsky1995, seifert1997configurations}, is that the physical behavior of a membrane on the length scale not much bigger than its own thickness, can be described with high accuracy by a purely geometric Hamiltonian \cite{canham1970minimum, helfrich1973elastic, evans1974bending}.
 Associated Euler-Lagrange equations \cite{zhong1987instability, zhong1989bending}, so called “shape equations”, are fourth order partial nonlinear differential equations, and finding a general analytical solution is typically difficult, even though it has been analytically \cite{0253-6102-59-2-14} and  numerically solved for some specific \cite{svetina1989membrane, seifert1990adhesion, lipowsky1991conformation, PhysRevA.44.1182, julicher1993domain, julicher1996shape, julicher1994shape, miao1994budding} and general cases \cite{heinrich1993nonaxisymmetric, kralj1993existence}.

 In fluid dynamics, material particles can be treated as a vertex of geometric figure and virtual layers as surfaces and equations of motion for such surfaces can be searched. We refer to the formalism as differentially variational surfaces (DVS) (or DVS formalism) \cite{2016arXiv160907765S}.

In this paper, we propose different approach to the 'shape choreography' problem. We use DVS formalism, tensor calculus of moving surfaces and the first law of thermodynamics to derive the final equation for the closed 2D surface dynamics 
(later on referred as surface) and to solve it analytically for the equilibrium case. In other words, we derive generic equations of motions for closed two-dimensional surfaces and without any \textit{a priori} symmetric assumptions, we show that constant mean curvature shapes are equilibrium solutions. In contrast to the Young-Laplace law these solutions, are universally correct descriptions of capillary surfaces as well as molecular surfaces. In addition, our equations of motions (\ref{38}-\ref{43}) are generic and exact. It advances our understanding of fluid dynamics because generalizes ideal magneto-hydrodynamic and Naiver-Stokes equations \cite{2016arXiv160907765S} and in contrast to Navier-Stokes, as we demonstrate in this paper, are trivially solvable for equilibrium shapes. To demonstrate the validity of these equations and their analytical solutions we apply them to micelles. Within our formalism it becomes simple task to show micelles lamellar, cylindrical, spherical shapes and assert their optimal spherical radius.  

For clarity, we shall give brief description of micelles and their structures.  A micelle consists of monolayer of lipid molecules containing hydrophilic head and hydrophobic tail. These amphiphilic molecules, in aqueous environment, aggregate spontaneously into a monomolecular layer held together due to a hydrophobic effect \cite{chandler2005interfaces, leikin1993hydration} (see also \cite{2016arXiv160907765S, svintradze2010hydrophobic, svintradze2015moving, svintradze2016cell, svintradze2017geometric}) by weak non-covalent forces \cite{tanford1973}. They form flexible surfaces that show variety of shapes of different topology, but remarkably in thermodynamic equilibrium conditions they are spherical, lamellar (plane) or cylindrical in shape.

\section{Methods}
In the section we provide basics of tensor calculus for moving surfaces and summarize the theorems we used directly or indirectly to derive equations for two-dimensional surface dynamics. Differential geometry preliminaries we used here are available in tensor calculus textbook \cite{grinfeld2010book} and in our work \cite{2016arXiv160907765S}.
\subsection{Basics of differential geometry.}

\begin{figure}[ht]
\includegraphics{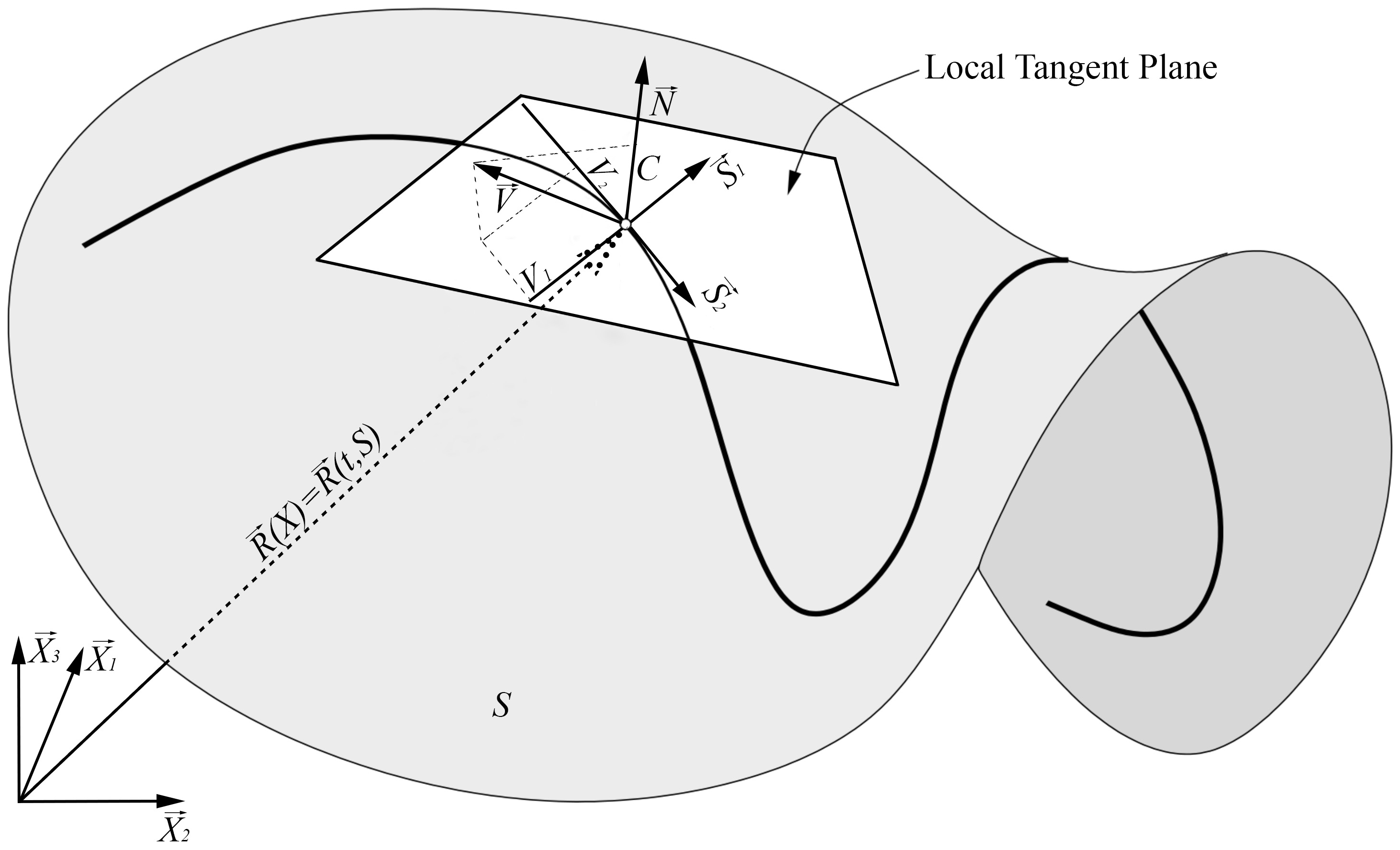}
\caption{\textbf{Graphical illustration of the arbitrary surface and its’ local tangent plane.} $\vec{S}_1,\vec{S}_2,\vec{N}$ are local tangent plane base vectors and local surface normal respectively. $\vec{X}_1,\vec{X}_2,\vec{X}_3$ are arbitrary base vectors of the ambient Euclidean space and $\vec{R}=\vec{R}(X)=\vec{R}(t,S)$ is radius vector of the point. $\vec{V}$ is arbitrary surface velocity and $C,V_1,V_2$ display projection of the velocity to the $\vec{N},\vec{S}_1,\vec{S}_2$ directions respectively.}
\label{fig:surface_velocity}
\end{figure}

Suppose that $S^i$ $(i=1,2)$ are the surface coordinates of the moving manifold (or the surface) $S$ and the ambient Euclidean space is referred to coordinates $X^\alpha$ (Figure \ref{fig:surface_velocity}).
Coordinates $S^i,X^\alpha$ are arbitrarily chosen so that sufficient differentiability is achieved in both, space and time. Surface equation in ambient coordinates can be written as $X^\alpha=X^\alpha(t,S^i)$. Let the position vector $\vec{R}$ be expressed in coordinates as 
\begin{equation} \label{19}
\vec{R}=\vec{R}(X^\alpha)=\vec{R}(t,S^i)
\end{equation}
Latin letters in indexes indicate surface related tensors. Greek letters in indexes show tensors related to Euclidean ambient space. All equations are fully tensorial and follow the Einstein summation convention. Covariant bases for the ambient space are introduced as $\vec{X}_\alpha=\partial_\alpha\vec{R}$, where $\partial_\alpha=\partial/\partial X^\alpha$. The covariant metric tensor is the dot product of covariant bases
\begin{equation} \label{20}
X_{\alpha\beta}=\vec{X}_\alpha \vec{X}_\beta
\end{equation}
The contravariant metric tensor is defined as the matrix inverse of the covariant metric tensor, so that $X^{\alpha\beta}X_{\beta\gamma}=\delta_\gamma^\alpha$, where $\delta_\gamma^\alpha$ is the Kronecker delta. As far as the ambient space is set to be Euclidean, the covariant bases are linearly independent, so that the square root of the metric tensor determinant is unit. Furthermore, the Christoffel symbols given by $\Gamma^\alpha_{\beta\gamma}=\vec{X}^\alpha\cdot\partial_\beta\vec{X}_\gamma$ vanish and set the equality between partial and curvilinear derivatives $\partial_\alpha=\nabla_\alpha$.

Now let's discuss tensors on the embedded surface with arbitrary coordinates $S^i$. Latin indexes throughout the text are used exclusively for curved surfaces and curvilinear derivative $\nabla_i$ is no longer the same as the partial derivative $\partial_i=\partial/\partial S^i$. Similar to the bases of ambient space, covariant bases of an embedded manifold are defined as $\vec{S}_i=\partial_i\vec{R}$ and the covariant surface metric tensor is the dot product of the covariant surface bases:
\begin{equation} \label{21}
S_{ij}=\vec{S}_i\cdot\vec{S}_j
\end{equation}
The contravariant metric tensor is the matrix inverse of the covariant one. The matrix inverse nature of covariant-contravariant metrics gives possibilities to raise and lower indexes of tensors defined on the manifold. The surface Christoffel symbols are given by $\Gamma^i_{jk}=\vec{S}^i\cdot\partial_j\vec{S}_k$ and along with Christoffel symbols of the ambient space provide all the necessary tools for covariant derivatives to be defined at tensors with mixed space/surface indexes:
\begin{align} \label{22}
\nabla_i T_{\beta k}^{\alpha j}=\partial_i T_{\beta k}^{\alpha j}+
X_i^\gamma\Gamma_{\gamma\nu}^\alpha T_{\beta k}^{\nu j}-
X_i^\gamma\Gamma_{\gamma\beta}^\mu T_{\mu k}^{\alpha j}  
+ \nonumber \\
\Gamma_{im}^jT_{\beta k}^{\alpha m}-\Gamma_{ik}^mT_{\beta m}^{\alpha j} 
\end{align}
where $X_i^\gamma$ is the shift tensor which reciprocally shifts space bases to surface bases, as well as space metric to surface metric; for instance, $\vec{S}_i=X^\alpha_i\vec{X}_\alpha$ and $S_{ij}=\vec{S}_i\cdot\vec{S}_j=X_i^\alpha\vec{X}_\alpha X^\beta_j\vec{X}_\beta=X_i^\alpha X^\beta_jX_{\alpha\beta}$. Note that in (\ref{22}) Christoffel symbols with Greek indexes are zeros.  

Using (\ref{20},\ref{22}), one may directly prove metrilinic property of the surface metric tensor $\nabla_iS_{mn}=0$, from where follows $\vec{S}_m\cdot\nabla_i\vec{S}_n=0$, meaning that $\vec{S}_m\bot\nabla_i\vec{S}_n$ are orthogonal vectors and as so $\nabla_i\vec{S}_n$ must be parallel to $\vec{N}$ the surface normal
\begin{equation} \label{23}
\nabla_i\vec{S}_j=\vec{N}B_{ij}
\end{equation} 
$\vec{N}$ is a surface normal vector with unit length and $B_{ij}$ is the tensorial coefficient of the relationship and is generally referred as the symmetric curvature tensor. The trace of the curvature tensor with upper and lower indexes is the mean curvature and its determinant is the Gaussian curvature. It is well known that a surface with constant Gaussian curvature is a sphere, consequently a sphere can be expressed as:
\begin{equation} \label{24}
B_i^i=\lambda
\end{equation}
where $\lambda$ is some non-zero constant. According to (\ref{23},\ref{24}), finding the curvature tensor defines the way of finding covariant derivatives of surface base vectors and as so, defines the way of finding surface base vectors which indirectly leads to the identification of the surface.

\subsection{Basics of tensor calculus for moving surfaces.}
All Equations written above are generally true for moving surfaces. We now turn to a brief review of definitions of coordinate velocity $V^\alpha$, interface velocity $C$ (which is the same as normal velocity), tangent velocity $V^i$ (Figure \ref{fig:surface_velocity}), time $\dot{\nabla}$-derivative of surface tensors and time differentiation of the surface integrals. The original definitions of time derivatives for moving surfaces were given in \cite{hadamard1968} and recently extended in tensor calculus textbook \cite{grinfeld2010book}.

Let's start from the definition of coordinate velocity $V^\alpha$ and show that the coordinate velocity is $\alpha$ component of the surface velocity. Indeed, by the definition
\begin{equation}
V^\alpha=\frac{\partial X^\alpha}{\partial t}
\end{equation}
On the other hand the position vector $\vec{R}$ given by (\ref{19}) is tracking the coordinate particle $S^i$. Taking into account partial time differential of (\ref{19}) and definition of ambient base vectors, we find
\begin{equation} \label{26}
\vec{V}=\frac{\partial\vec{R}(t,S^i)}{\partial t}=\frac{\partial\vec{R}}{\partial X^\alpha}\frac{\partial X^\alpha(t,S^i)}{\partial t}=V^\alpha\vec{X}_\alpha
\end{equation}
Therefore, $V^\alpha$ is ambient component of the surface velocity $\vec{V}$. Taking into account (\ref{26}), normal component of the surface velocity is dot product with the surface normal, so that
\begin{equation} \label{27}
C=\vec{V}\cdot\vec{N}=V_\alpha\vec{X}^\alpha N^\beta\vec{X}_\beta=V_\alpha N^\beta\delta_\beta^\alpha=V_\alpha N^\alpha
\end{equation}
It is easy to show that the normal component $C$ of the coordinate velocity, generally referred as interface velocity, is invariant in contrast with coordinate velocity $V^\alpha$ and its sign depends on a choice of the normal. The projection of the surface velocity on the tangent space (Figure \ref{fig:surface_velocity}) is tangential velocity and can be expressed as
\begin{equation}\label{28}
V^i=V^\alpha X_\alpha^i
\end{equation}
Taking (\ref{27},\ref{28}) into account one may write surface velocity as $\vec{V}=C\vec{N}+V^i\vec{S}_i$. Graphical illustrations of coordinate velocity $V^\alpha$, interface velocity $C$ and tangential velocity $V^i$ are given on Figure \ref{fig:surface_velocity}. There is a clear geometric interpretation of the interface velocity \cite{grinfeld2010book}. Let the surfaces at two nearby moments of time $t$ and $t+\Delta t$ be $S_t$, $S_{t+\Delta t}$ correspondingly. Suppose that $A\in S_t$ (point on $S_t$) and the corresponding point $B\in S_{t+\Delta t}$, $B$ has the same surface coordinates as $A$ (Figure \ref{fig:surface_derivative}), then $\vec{AB}\approx\vec{V}\Delta t$. Let $P$ be the point, where the unit normal $\vec{N}\in S_t$ intersect the surface $S_{t+\Delta t}$, then for small enough $\Delta t$, the angle $\angle APB\rightarrow\pi/2$ and $AP\rightarrow\vec{V}\cdot\vec{N}\Delta t$, therefore, $C$ can be defined as
\begin{equation}
C=\lim_{\Delta t\rightarrow 0}\frac{AP}{\Delta t}
\end{equation}
and can be interpreted as the instantaneous velocity of the interface in the normal direction. It is worth of mentioning that the sign of the interface velocity depends on the choice of the normal. Although $C$ is a scalar, it is called interface velocity because the normal direction is implied.

\subsection{Invariant time differentiation.}
Among the key definitions in calculus for moving surfaces, perhaps one of the most important is the invariant time derivative $\dot\nabla$. As we have already stated, invariant time derivative is already well defined in the literature \cite{hadamard1968, grinfeld2010book}. In this paragraph, we just give geometrically intuitive definition.

\begin{figure}[ht]
\includegraphics{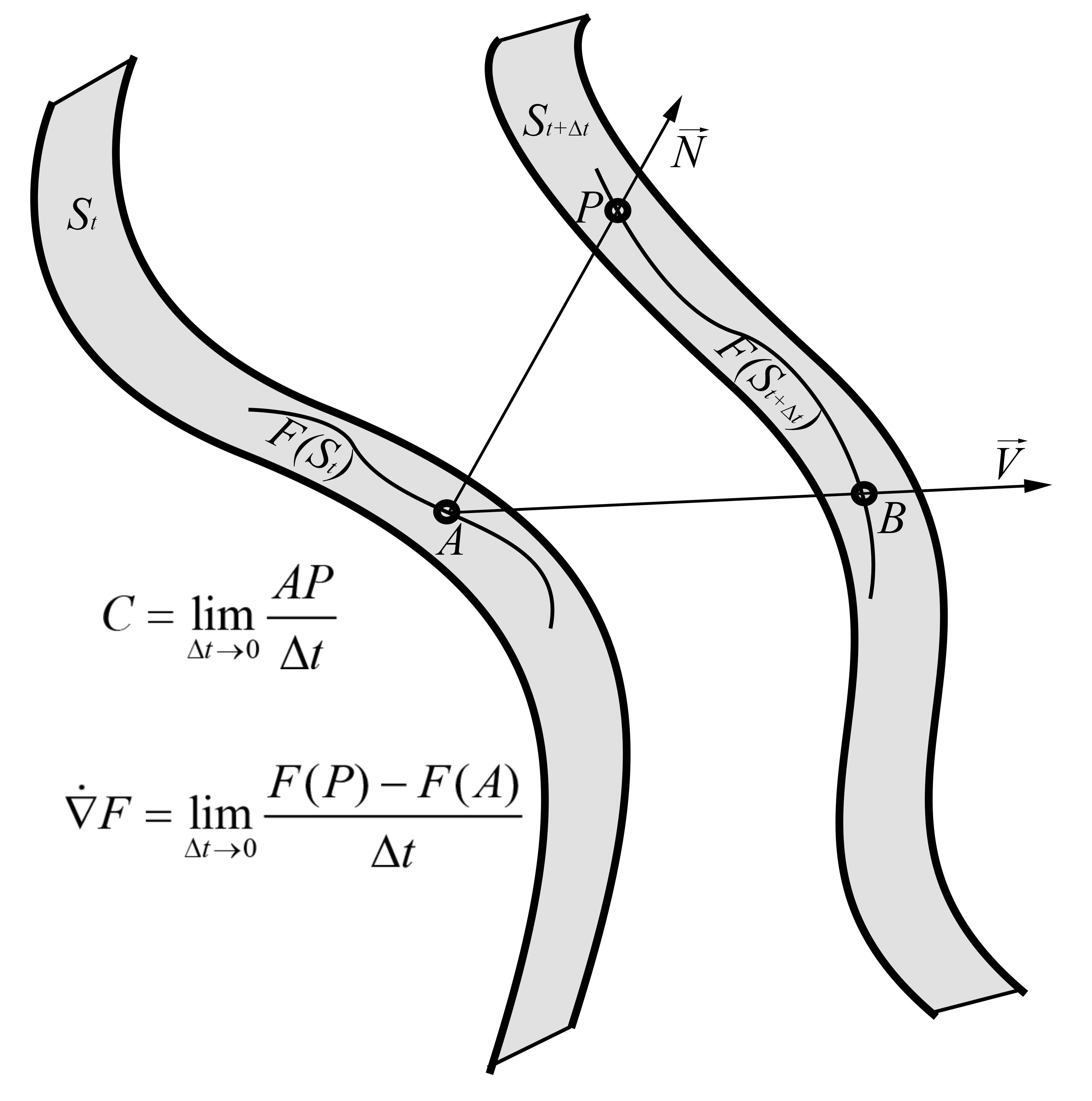}
\caption{\textbf{Geometric interpretation of the interface velocity $C$ and of the curvilinear time derivative $\dot\nabla$ applied to invariant field $F$.} $A$ is arbitrary chosen point so that it lays on $F(S_t)\in S_t$ curve and $B$ is its' corresponding point on the $S_{t+\Delta t}$ surface. $P$ is the point where $S_t$ surface normal, applied on the point $A$, intersects the surface $S_{t+\Delta t}$. By the geometric construction, for small enough $\Delta t\rightarrow 0$, $\angle APB\rightarrow\pi/2$, $\vec{AB}\approx\vec{V}\Delta t$ and $AP\approx\vec{V}\vec{N}\Delta t$. On other hand, by the same geometric construction the field $F$ in the point $B$ can be estimated as $F(B)\approx F(A)+\Delta t\partial F/\partial t$, while from viewpoint of the $S_{t+\Delta t}$ surface the $F(B)$ value can be estimated as $F(P)+\Delta tV^i\nabla_iF$, where $\nabla_iF$ shows rate of change in $F$ along the surface $S_{t+\Delta t}$ and along the directed distance $BP\approx\Delta tV^i$.}
\label{fig:surface_derivative}
\end{figure}

Suppose that invariant field $F$ is defined on the surface at all time. The idea behind the invariant time derivative is to capture the rate of change of $F$ in the normal direction. Physical explanation of why the deformations along the normal direction are so important, we give below in integration section. This is similar to how $C$ measures the rate of deformation in the normal direction. Let for a given point $A\in S_t$, find the points $B\in S_{t+\Delta t}$ and $P$ the intersection of $S_{t+\Delta t}$ and the straight line orthogonal to $S_t$ (Figure \ref{fig:surface_derivative}). Then, the geometrically intuitive definition dictates that
\begin{equation} \label{30}
\dot\nabla F=\lim_{\Delta t\rightarrow 0}\frac{F(P)-F(A)}{\Delta t}
\end{equation}
As far as (\ref{30}) is entirely geometric, it must be an invariant (free from choice of a reference frame). From the geometric construction one can estimate value of $F$ in point $B$, so that
\begin{equation} \label{31}
F(B)\approx F(A)+\Delta t\frac{\partial F}{\partial t}
\end{equation}
On the other hand, $F(B)$ is related to $F(P)$ because $B,P\in S_{t+\Delta t}$
and are nearby points on the surface $S_{t+\Delta t}$, then according to definition of covariant derivative
\begin{equation} \label{32}
F(B)\approx F(P)+\Delta tV^i\nabla_iF
\end{equation}
since $\nabla_iF$ shows rate of change in $F$ along the surface and $\Delta t \cdot V^i$ captures the directed distance $BP$. Determining $F(A),F(P)$ values from (\ref{31},\ref{32}) and putting it in (\ref{30}), gives
\begin{equation}\label{33}
\dot\nabla F=\frac{\partial F}{\partial t}-V^i\nabla_i F
\end{equation}
Extension of the definition (\ref{33}) for any arbitrary tensors with mixed space and surface indexes is given by the formula
\begin{align} \label{34}
\dot{\nabla}T_{\beta j}^{\alpha i}= \frac{\partial T_{\beta j}^{\alpha i}}{\partial t}-V^k\nabla_kT_{\beta j}^{\alpha i}+V^\gamma\Gamma_{\gamma\mu}^\alpha T_{\beta j}^{\mu i}-V^\gamma\Gamma_{\gamma\beta}^\mu T_{\mu j}^{\alpha i} \nonumber \\
+\dot{\Gamma}_k^iT_{\beta j}^{\alpha k}-\dot{\Gamma}_j^kT_{\beta k}^{\alpha i}
\end{align}
The derivative commutes with contraction, satisfies sum, product and chain rules, is metrinilic with respect to the ambient metrics and does not commute with the surface derivative \cite{grinfeld2010book}. Also from (\ref{30}) it is clear that the invariant time derivative applied to time independent scalar
vanishes. Christoffel symbol $\dot\Gamma^i_j$ for moving surfaces is defined by the formula $\dot\Gamma^i_j=\nabla_jV^i-CB^i_j$.

\subsection{Time differentiation of integrals.}
The remarkable usefulness of the calculus of moving surfaces becomes evident from two fundamental formulas for integrations that govern the rates of change of volume and surface integrals due to the deformation of the domain \cite{grinfeld2010book}. For instance, in evaluation of the least action principle of the Lagrangian there is a central role for time differentiation of the surface and space integrals, from where the geometry dependence is rigorously clarified.

For any scalar field $F=F(t,S)$ defined on a Euclidean domain $\Omega$ with boundary $S$ evolving with the interface velocity $C$, the evolution of the space integral and surface integral for closed surfaces are given by the formulas
\begin{align}
\frac{d}{dt}\int_\Omega Fd\Omega=&\int_\Omega\frac{\partial F}{\partial t}d\Omega+\int_SCFdS \label{35} \\
\frac{d}{dt}\int_S FdS=&\int_S\dot\nabla FdS-\int_SCFB_i^idS \label{36}
\end{align}
The first term in the integral represents the rate of change of the tensor field, while the second term shows changes in the geometry. Of course there are rigorous mathematical proofs of these formulas in the tensor calculus textbooks. We are not going to reproduce proof of these theorems here, but instead we give less rigorous but completely intuitive explanation of why only interface velocity has to be count. Rigorous mathematical proof follows from fundamental theorem of calculus
\begin{equation}
\frac{d}{dt}\int_a^{b(t)}F(t,x)dx=\int_a^{b(t)}\frac{\partial F(t,x)}{\partial t}dx+b'(t)F(t,b(t))
\end{equation}
In the case of volume integral or surface integral it can be shown that $b'(t)$ is replaced by interface velocity $C$. 

Intuitive explanation is pretty simple. Propose there is no interface velocity then closed surface velocity only has tangent component. For each given time tangent velocity (if there is no interface velocity) translates each point to its neighboring point and therefore, does not add new area to the closed surface (or new volume to the closed space, or new length to the closed curve). As so, tangential velocity just induces rotational movement (or uniform translational motion) of the object and can be excluded from additive terms in the integration. Perhaps, it is easier to understand this statement for one dimensional motion. Let's assume that material point is moving along some trajectory (some closed curve or loop), then, in each point, the velocity of the material point is tangential to the curve. Now one can translate this motion into the motion of the closed curve where the loop has only tangential velocity. In this aspect, the embedded loop only rotates (uniformly translates in the plane) without changing the length locally, therefore tangential velocity of the curve does not add new length to the curve (Same is true for open curve with fixed ends).

\section{General equations of surface motions}
Fully non-restrained and exact equations for moving three-dimensional surfaces in electromagnetic field, when the interaction with an ambient environment is ignored, reads
\begin{widetext}
\begin{align}
\dot\nabla\rho+\nabla_i(\rho V^i)=&\rho CB^i_i \label{38}\\
\partial_\alpha(\rho V^\alpha(\dot\nabla C+2V^i\nabla_iC+V^iV^jB_{ij})-V^\alpha(\frac{1}{4\mu_0}F_{\mu\nu}F^{\mu\nu}+A_\mu J^\mu))=&fa \label{39} \\
\int_S\rho V_i(\dot\nabla V^i+V^j\nabla_jV^i-C\nabla^iC-CV^jB_j^i)dS=&\int_\Omega f^ia_id\Omega \label{40}
\end{align}
\end{widetext}
where $\rho$ is the surface mass density, $V^\alpha$,  $V^i$  are coordinate and tangential components of the surface velocity, $C$ is interface velocity, $\alpha=0,1,2,3$ for Minkowski four-dimensional space-time ambient space, $i=0,1,2$ for pseudo-Riemannian manifold (surface), $B_{ij}$ is the surface curvature tensor, $F^{\mu\nu}$ is electromagnetic tensor, $F^\alpha=J^\alpha-\partial_\beta F^{\beta\alpha}$, $J^\alpha$ is $\alpha$ component of $\vec{J}=(J^\alpha)$ four current, $f,f^i$ are normal and tangential components of $\vec{F}=(F^\alpha)$, $a$, $a_i$ are the
normal and tangential components of the partial time derivative of the four vector potential $\vec{A}=(A_\alpha)$, $S$, $\Omega$ stand for surface and space integrals respectively. Exact derivation of (\ref{38}-\ref{40}) is given in our work \cite{2016arXiv160907765S}, we don't reproduce derivation of this set in this paper, rather just mention that first one is the consequence of mass conservation, second and third equations come from minimum action principle of a Lagrangian and imply motion in normal direction (\ref{39}) and in tangent direction (\ref{40}). 

 For two dimensional surface dynamics, Minkowskian space becomes Euclidean, so that $\alpha=1,2,3$ and the surface is two-dimensional Riemannian manifold $i=1,2$. 
 So that, after modeling the potential energy as a negative volume integral of the internal pressure and inclusion interaction with an environment, (\ref{38}-\ref{40}) further simplifies as
\begin{widetext}
\begin{align}
\dot\nabla\rho+\nabla_i(\rho V^i)=&\rho CB^i_i \label{41} \\
\partial_\alpha(\rho V^\alpha(\dot\nabla C+2V^i\nabla_iC+V^iV^jB_{ij})+V^\alpha(P^++\Pi))=&-V^\alpha\partial_\alpha(P^++\Pi) \label{42} \\
\rho V_i(\dot\nabla V^i+V^j\nabla_jV^i-C\nabla^iC-CV^jB_j^i)=&0 \label{43}
\end{align}
\end{widetext}
where $P^+$, $\Pi$ are internal hydrodynamic and osmotic pressures, respectively. Derivation of (\ref{38}-\ref{40}) can be found in \cite{2016arXiv160907765S}. We derive (\ref{41}-\ref{43}) in appendix section. It is noteworthy that from the last equations set only the second equation (\ref{42}) differs from the dynamic fluid film equations  \cite{grinfeld2009exact, grinfeld2010book}
\begin{equation}
\rho(\dot\nabla C+2V^i\nabla_iC+V^iV^jB_{ij})=\sigma B_i^i \label{44}
\end{equation}
where $\sigma$ is surface tension. (\ref{44}) is only valid when the surface can be described with time invariable surface tension \cite{grinfeld2009exact, grinfeld2010book}, meaning that the surface is homogeneous and the surface tension is constant, while (\ref{42}) does not have that restriction. Using (\ref{44}) in (\ref{42}) and taking into account that in equilibrium processes internal pressure is the same as external pressure, one gets exactly the same equation of motion in normal direction (\ref{10}) as we get from using the first law of thermodynamics (see below).
\begin{equation}\label{45}
\partial_\alpha(\sigma V^\alpha B_i^i+(P^++\Pi)V^\alpha)=-(\partial_\alpha P^++\partial_\alpha\Pi)V^\alpha
\end{equation}
It is worth of mentioning that (\ref{41}-\ref{43}) also follows from (\ref{38}-\ref{40}) if one applies same formalism as it is given in (\ref{1}-\ref{3}). Indeed, for relatively slowly moving surfaces space is three dimensional Euclidean so that $\alpha=1,2,3$, the surface is two-dimensional Riemannian ($i=1,2$) and the potential energy becomes
\begin{align}\label{46}
U=&\int_\Omega(\frac{1}{4\mu_0}F_{\mu\nu}F^{\mu\nu}+A_\mu J^\mu) \nonumber \\
=&\int_\Omega(-\frac{\epsilon_0}{2}E^2+\frac{1}{\mu_0}B^2-q\varphi+\vec{A}\vec{J})d\Omega
\end{align}
where $\vec{E},\vec{B}$ are electric and magnetic fields and $q,\varphi,\vec{A},\vec{J}$ are charge density, electric potential, magnetic vector potential and current density vector respectively. Using (\ref{1}-\ref{3}) formalism into account, we find
\begin{equation}\label{47}
dU=-(P^++\Pi)d\Omega=(-\frac{\epsilon_0}{2}E^2+\frac{1}{\mu_0}B^2-q\varphi+\vec{A}\vec{J})d\Omega
\end{equation}
Taking into account (\ref{47}) and that the pressure comes from the normal force applied to the surface, we find ${fa=-V^\alpha\partial_\alpha(P^++\Pi)}$ and in tangent direction $f^ia_i=0$, then (\ref{38}-\ref{40}) becomes (\ref{41}-\ref{43}). Electromagnetic potential energy can be generalized if one takes into account environment, which enters in energy terms as bound and free charges and electric/magnetic fields are replaced by polarization and magnetization vectors \cite{2016arXiv160907765S}.

\section{Results and Discussion}
\subsection{General assumptions.}

\begin{figure}[ht]
	\includegraphics{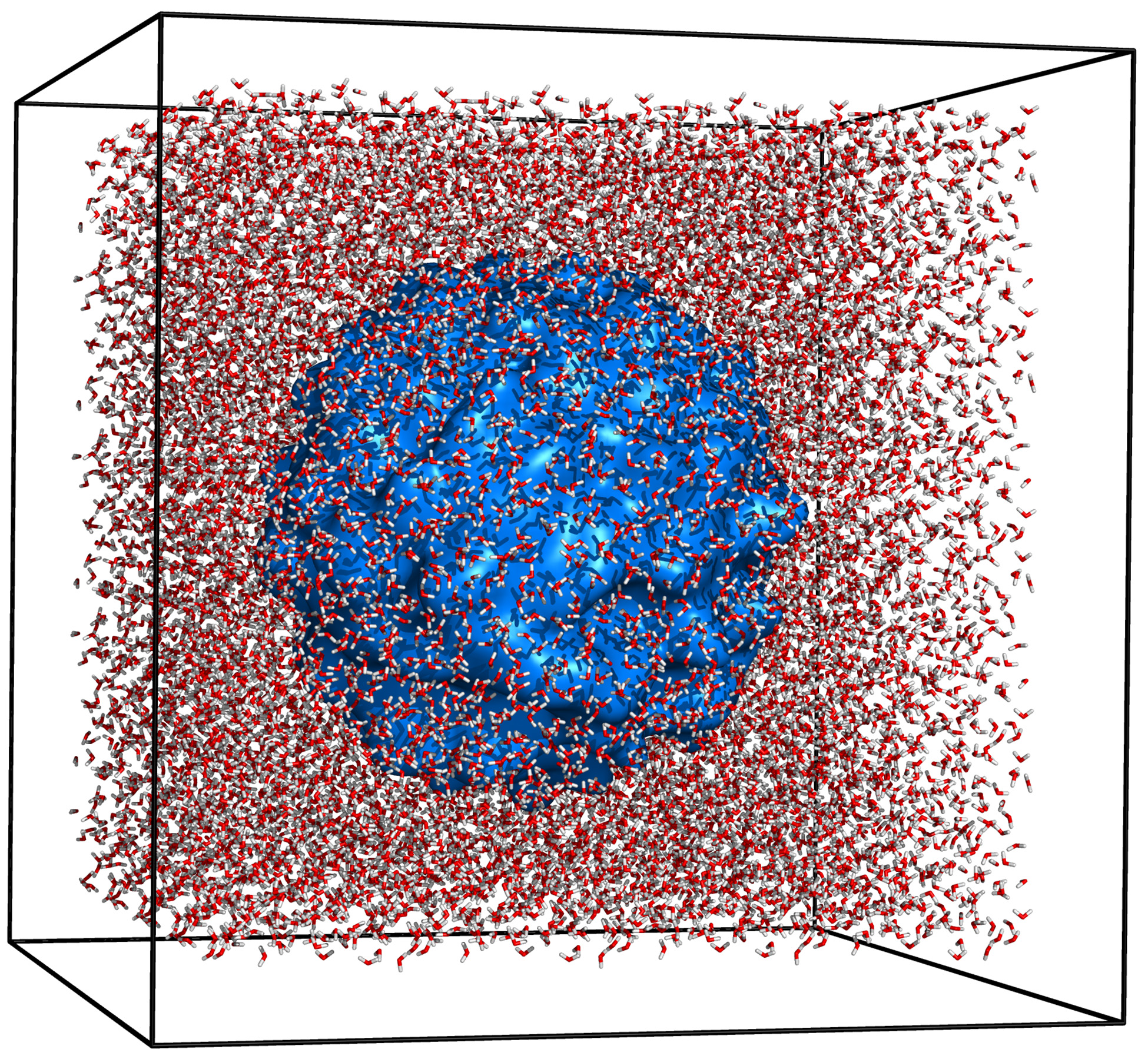}
	\caption{\textbf{Graphical illustration of the isolated system containing aqueous solution.}  Water molecules are represented as red and white sticks. The system boundary is shown as white faces with black edges. The subsystem-micelle is closed surface, blue blob in the center of the system.}
	\label{fig:Micelle}
\end{figure}

In this section we apply basics of thermodynamics and fundamental theorems of calculus of moving surfaces to demonstrate shortest derivation of the equation, describing motion of homogeneous, closed two dimensional surface with time invariable surface tension at normal direction (\ref{45}). We consider the system consisted of aqueous media with the formed closed surface in it  (Figure \ref{fig:Micelle}). The system is isolated with constant temperature and there is no absorbed or dissipated heat on the surface; in other words, a process is adiabatic. 
 According to the first law of thermodynamic, as far as there is no dissipated or absorbed heat, the change of the internal energy of the surface must be
\begin{equation}
dE=\delta W \label{1}
\end{equation}
where $\delta W$ is infinitesimal work done on the subsystem and $dE$ is infinitesimal change of the internal energy. Because the temperature of the system is constant, the differential of the subsystems' internal energy can be remodeled as
\begin{equation}
dE=dU \label{2}
\end{equation}
where $U$ is the total potential energy of the surface. By the definition the elementary work done on the subsystem is
\begin{equation}
\delta W=(P^-+\Pi)d\Omega \label{3}
\end{equation}
where , $P^-,\Pi$ are external hydrodynamic and osmotic pressures applied on the surface by the surroundings correspondingly and $\Omega$ is the volume that surface encloses with boundary of $S$ surface area. Let's propose that the surface is homogeneous (i.e material particles are homogeneously distributed on the surface) so that the total potential energy is integration of the potential energy per unit area over the surface, then
\begin{equation}
dU=\sigma dS \label{4}
\end{equation}
where $\sigma$ is the potential energy per unit area and is called surface tension in the paper. As far as we discuss simplest case of the system consisted of aqueous medium and single closed surface, we can suggest that the surface tension is not time variable. Using (\ref{1}-\ref{4}) after few lines of algebra, we fined 
\begin{equation}
\int_S \sigma dS=\int_\Omega (P^-+\Pi)d\Omega \label{5}
\end{equation}
Assuming the surface is moving so that (\ref{5}) stays valid for any time variations, then time differentiation of the left side must be equal to time differentiation of right integral. As far as on the right hand side we have space integral, time differentiation can be taken into the integral, using general theorems for differentiation of space and surface integrals (\ref{35}-\ref{36}), so that integration theorem for space integral holds and the convective and advective terms due to volume motion are considered
\begin{align} \label{6}
\frac{d}{dt}\int_\Omega (P^-+\Pi)d\Omega=\int_\Omega (\partial_\alpha P^-+\partial_\alpha\Pi)\frac{\partial X^\alpha}{\partial t}d\Omega \nonumber \\
+\int_SC(P^-+\Pi)dS
\end{align}
To calculate time derivative of the surface integral we have to take into account the theorem about time differentiation of the surface integral (\ref{36}), from which follows that for time invariable surface tension
\begin{equation} \label{7}
\frac{d}{dt}\int_S\sigma dS=\int_S-\sigma CB_i^idS
\end{equation}
Where $C=V^\alpha N_\alpha$ is interface velocity, $N_\alpha$ is $\alpha$ component of the surface normal and ${V=\partial X^\alpha/\partial t}$ is coordinate velocity, $X^\alpha$ is general coordinate and $B_i^i$ is the trace of the mixed curvature tensor generally known as mean curvature. After few lines of algebra putting (\ref{5}-\ref{7}) together, we find
\begin{equation} \label{8}
\int_S(\sigma CB_i^i+C(P^-+\Pi))dS=-\int_\Omega(\partial_\alpha P^-+\partial_\alpha\Pi)V^\alpha d\Omega
\end{equation}
Generalized Gauss theorem converts the surface integral of the left hand side of (\ref{8}) into space integral, so that
\begin{align} \label{9}
\int_SN_\alpha V^\alpha(\sigma B_i^i+P^-+\Pi)dS=\int_\Omega\partial_\alpha(\sigma V^\alpha B_i^i \nonumber \\
+(P^-+\Pi)V^\alpha)d\Omega
\end{align}
Combination of (\ref{8}) and (\ref{9}) immediately gives equation of motion for surface in normal direction
\begin{equation}\label{10}
\partial_\alpha(\sigma V^\alpha B_i^i+(P^-+\Pi)V^\alpha)=-(\partial_\alpha P^-+\partial_\alpha\Pi)V^\alpha
\end{equation}
For equilibrium processes internal and external pressures are identical ${P^-=P^+}$, so that (\ref{10}) becomes identical to the equation of motion in normal direction observed from master equations (\ref{41}-\ref{45}). Also, we should note that (\ref{10}) is only valid for motion of the homogeneous surfaces with time invariable surface tension at normal direction, therefore, it does not display any deformation in tangent directions. (\ref{10}) further simplifies when the surface comes in equilibrium with the solvent where divergence of the surface velocity $\partial_\alpha V^\alpha$ (stationary interface) along with $\partial P/\partial t$ (where $P=P^-+\Pi$) vanishes, then the solution to (\ref{10}), taking into account the condition (\ref{6}), becomes
\begin{equation} \label{11}
B^i_i=-\frac{P}{\sigma}
\end{equation}
The result (\ref{11}) shows that the solution is constant mean curvatures (CMC) surfaces. Such CMC are rare and can be many if one relaxes the condition we restricted to the system. We consider isolated system where the surface is closed subsystem, these two preconditions mathematically mean that the surface we discuss is compact embedded surface in $\mathbb{R}^3$. According to A. D. Alexandrov uniqueness theorem for surfaces, a compact embedded surface in $\mathbb{R}^3$ with constant non-zero mean curvature is a sphere \cite{alexandrov1958}. Correspondingly the solution (\ref{11}) is a sphere (as far as we have compact two-manifold in the Euclidean space). When
\begin{equation}
\frac{P}{\sigma}\neq 0
\end{equation}
the surface is spheroid (or a cylinder if one relaxes compactness restriction making the cylinder infinitely long) and becomes plane (again when compactness argument is relaxed) or other zero mean curvature shape when compactness argument is not relaxed but contour of the surface remains fixed. This surprisingly simple and elegant derivation explains all the shapes surfaces can adopt in aqueous solution at equilibrium conditions.\footnote{Even though we set environment as aqueous, it enters into equations as osmotic pressure term, which due to a generality of arguments can be anything. Therefore, as a medium one may pick any liquid or gas.} If the compactness condition is relaxed then (\ref{11}) predicts that in addition to cylinder and plane all other CMC surfaces are also equilibrium shapes for moving surfaces. Taking into account that the surface tension in general can be a function of many
variables, such as Gaussian curvature, bending rigidity, spontaneous curvature, molecules concentration, geometry of surfactant molecules and etc., then (\ref{10}) may predict possible deformations of differently shaped surfaces and their wide range of static shapes. In fact, if considered that surface tension, which is defined as potential energy per unit area, can be a function of mean curvature $\sigma=\sigma(B^i_i)$, then Taylor expansion of $\sigma(B^i_i)$ naturally rises all additional terms. These generalizations and temperature fluctuations can be included in the equations, but it is not scope of this paper and should be addressed separately. One may even propose $\sigma$ as time independent the Helfrich Hamiltonian and then (\ref{10}, \ref{11}) will become equation of static shapes for homogeneous surfaces with time invariable surface tension.

\subsection{Physical application, micelle.}

We can put equation (\ref{10}) and its solution (\ref{11}) under the test for homogeneous  micellar surface equilibrated with the aqueous solution. Based on (\ref{11}) we can calculate minimal value of a micelle radius. The value of the trace of the mixed curvature tensor for a sphere is
\begin{equation} \label{13}
B^i_i=-\frac{2}{R}
\end{equation}
where $R$ is radius. 
\begin{figure}[ht]
	\includegraphics{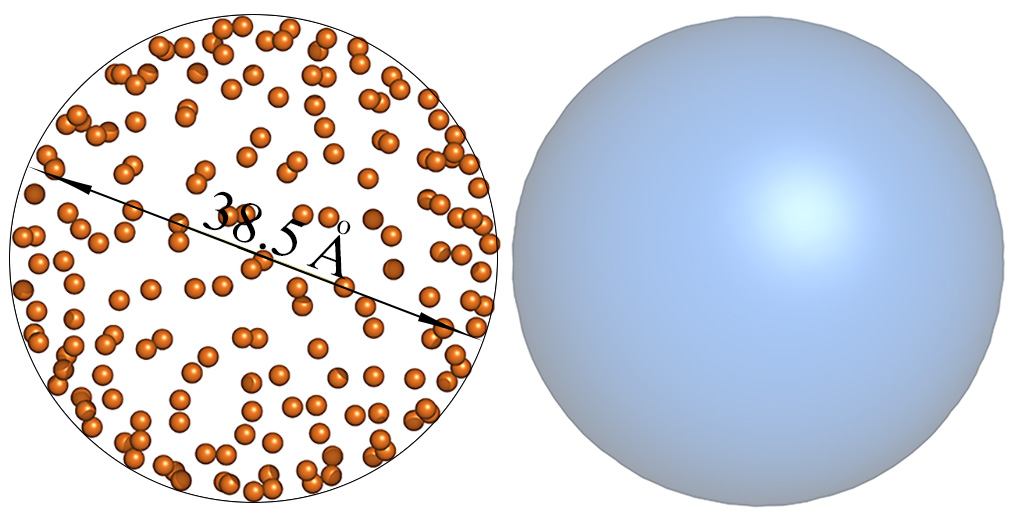}
	\caption{\textbf{ Simulated three dimensional coordinates of the micelle in aqueous solution display sphere with diameter $38.5{\AA}$.} (Left) dihexanoylphosphatidylcholine (DHPC phospholipids) are modeled as orange balls. (Right) Gaussian mapping at contour resolution $8{\AA}$ of the micelle shows spherical structure.}
	\label{fig:Simulated_micelle}
\end{figure}

Let's calculate value of the surface pressure when the micelle still can exist. Lipids in a micelle are confined in the surface by hydrophobic interactions with average energy in the range of hydrogen bonding. As far as values of hydrogen bonding energy are somewhat uncertain in the literature, by the first approximation we take average energy for the hydrogen bonding energy interval and assign it to the lipid molecule.
Low boundary of the interval (minimum energy) for $XH\cdot\cdot\cdot Y$ hydrogen bond is about 1 kJ per mol ($CH\cdot\cdot\cdot C$ unit) and high boundary is about 161 kJ per mol ($FH\cdot\cdot\cdot F$ unit), the low and high values are taken according to references \cite{larson1984gas, emsley1980very}. Therefore, average energy is about $(1+161)/2=81kJ/mol\approx 13\cdot 10^{-20}J$. To estimate hydrogen bonding energy per molecule with the undefined shape (lipid molecule) in the first approximation is to assign average energy to it and consider the spherical shape with the gyration radius. Of course it is low level approximation, but even such rough calculations produce reasonable results. After all these rough estimations the pressure to move one lipid from the surface, in order to induce critical
deformations of the surface, is about average energy per the average volume of the lipid molecule
\begin{equation} \label{14}
P\approx \frac{3\cdot 13\cdot 10^{-20}}{4\pi r_G^3}\approx 3.1\cdot 10^7N/m^2
\end{equation}
where $4\pi r_G^3/3$ is the estimated volume of a lipid molecule considered as sphere with the gyration radius $r_G\approx 1nm$. On the other hand, surface tension of a fluid monolayer at optimal packing of the lipids is about $\sigma\approx 3\cdot 10^{-2}N/m$ \cite{jahnig1996surface, Israelachvili1977, israelachvili2011intermolecular}, using these and (\ref{13},\ref{14}) in (\ref{11}) the estimated micelle radius is
\begin{equation} \label{15}
R\approx\frac{2\cdot 3\cdot 10^{-2}}{3.1\cdot 10^7}=19.3\pm 0.1{\AA}
\end{equation}

\begin{figure*}
	\includegraphics[width={\textwidth}, height=8cm]{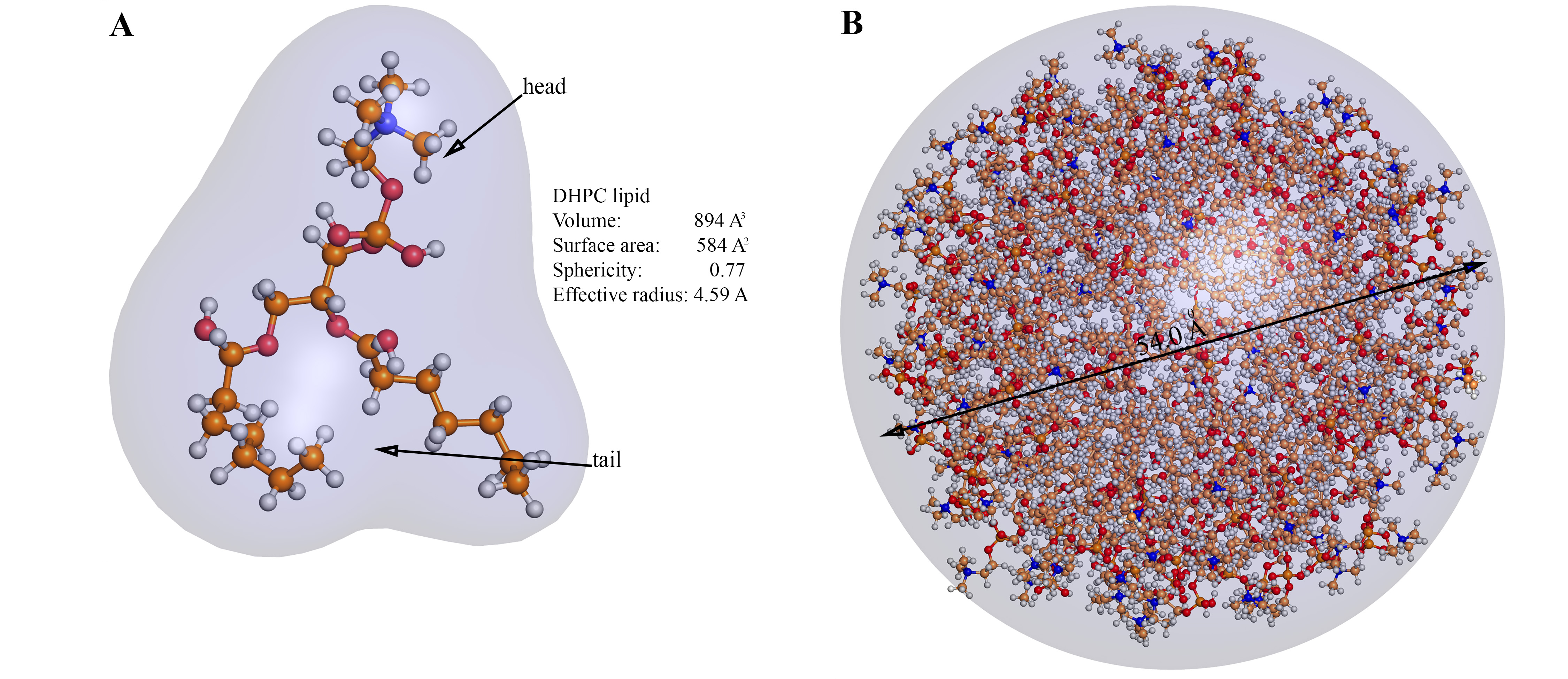}
	\caption{\textbf{All atom simulation of DHPC micelle.} (A) The figure shows a geometry of the DHPC surfactant molecule used in simulation and gives parametric description of volume, surface area, sphericity and effective radius. (B) Indicates atomistic simulation result contoured by Gaussian map and the diameter of the micelle, measured by PyMol. The diameter of the simulated micelle appears to be $54.0{\AA}$ with the uncertainty of the measurement $0.1{\AA}$.}
	\label{fig:atomistic_model}
\end{figure*}

These calculations put the minimum radius in nanometer scale and is in very good agreement with experimental as well as computational frameworks \cite{feller1995computer, egberts1994molecular}. To further validate the (\ref{15}) result, we ran a CHARMM based Micelle Builder simulation \cite{jo2008charmm, cheng2013charmm} for 100 phospholipid molecules (DHPC lipids). The simulation result (Figure \ref{fig:Simulated_micelle}) generated a spherical micelle with diameter $38.5\pm0.1{\AA}$. These calculations indeed indicate that even such rough estimations produce reasonable accuracy.

To get more convincing estimations it is necessary to take into account that neither lipids are spherical nor hydrophobic interactions per lipid are average energy of single hydrogen bond. In the second approximation lipids are no longer undefined spheres, but have well defined surfactant geometry. The Hydrophobic energy is no longer average energy of single hydrogen bond, but is 1 kJ per mol per $-CH_2-$ unit. In all atom simulations we used dihexanoylphosphatidylcholine (DHPC) lipid molecule having $12-CH_2-$ units (Figure \ref{fig:atomistic_model}) per hydrophobic tail, so hydrophobic energy is about $12kJ/mol\approx 1.99\cdot 10^{-20}J$. Accurate calculation of the lipid molecule volume using cavity, channel and cleft volume calculator \cite{voss20103v}, gives the volume estimation of about $894{\AA}^3$. Using this value, one gets
\begin{equation}
P\approx\frac{1.99\cdot 10^{-20}}{0.894\cdot 10^{-27}}\approx2.22\cdot10^7N/m^2
\end{equation}
On other hand, using the same surface tension of a fluid monolayer at the optimal packing of the lipids, one gets $R=27\pm0.1{\AA}$. All atom simulation also generates spherical structure with diameter $54\pm0.1{\AA}$ (Figure \ref{fig:atomistic_model}). There is still some uncertainty in this estimation because we assigned 1 kJ/mol energy per $-CH_2-$ unit and we based on references data \cite{larson1984gas, emsley1980very}, while in other literature it is mentioned that the hydrophobic interactions are about 4 kJ/mol per $-CH_2-$ unit \cite{leitmannova2011}. In our opinion, this discrepancy can be resolved if one calculates hydrophobic energy based on the potential energy
\begin{equation} \label{17}
U=-\int_\Omega\frac{\epsilon_0}{2}E_{CH_2}^2d\Omega
\end{equation}
where $\vec{E}_{CH_2}$ is electric field per $-CH_2-$, $\epsilon_0$ is dielectric constant in the vacuum and $\Omega$ stands for the volume of the lipid molecules. (\ref{17}) directly emerges from $F_{\mu\nu}F^{\mu\nu}$ term written in the equations of motion (\ref{38}-\ref{40} and \ref{46}-\ref{47}). For electrostatics
\begin{equation}
U=\int_\Omega\frac{1}{4\mu_0}F_{\mu\nu}F^{\mu\nu}d\Omega=-\int_\Omega\frac{\epsilon_0}{2}E_{CH_2}^2d\Omega
\end{equation}
so one should go to the scrutiny of calculating electric field for each $-CH_2-$ units, then take a sum of the electric field and square it (we are not going to do it in this paper). Also, one may ask why the hydrophilic interaction energy is not taken into account in these calculations.
Hydrophilic head of the lipid molecule is in contact with water molecules so there is no work needed to drag it in aqueous solution from the lipids layer. Therefore, hydrophilic interaction energy can be neglected. The most work goes on overcoming hydrophobic interactions between lipid tails.

\section{Conclusion}

We have presented a framework for the analysis of two dimensional surface dynamics (identified as micelle in the text) using first law of thermodynamics and calculus of moving surfaces. In final equations of normal motion (\ref{10},\ref{45}) we assume that a surface is homogeneous and has time invariable surface tension. However, the general equations (\ref{41}-\ref{43}) don not have these constrains and indicate arbitrary motion along normal deformation, as well as into tangent directions, but are analytically more complex. The solution to the normal equations of motion in equilibrium conditions are surprisingly simple and display all possible equilibrium shapes. We applied the formalism to estimate micelle optimal radius and compared estimations to all atom simulations. Even for low-level approximations, we found remarkable agreement between theoretically calculated radius and one obtained from atomistic simulations and from experiments. One can  readily apply the theory to any closed surfaces; such are vesicles, membranes, water droplets or soap films.

As a final remark, even though the analytic solution (\ref{11}) looks like generalized Young-Laplace law, the difference is obvious. $B_i^i$ is a trace of mixed curvature tensor, known as mean curvature, and when the mean curvature is constant, it defines whole class of constant mean curvature (CMC) surfaces. Generalized the Young-Laplace law is \textit{a priori} formulated for spherical morphologies and therefore in some particular cases can be obtained from (\ref{11}) constant mean curvature shapes. The condition for holding the particular case is a compactness. However, the compactness argument can be relaxed in our derivation if the considered system is set to be much larger than the subsystem. Therefore, the solution (\ref{11}) effectively predicts formation of all CMC surfaces while Young-Laplace law is correct for spherical structures alone. Also in derivation of Young-Laplace relation one of cornerstone idea is suggestion of spherical symmetries, while our derivation is free of symmetries and explains why CMC surfaces are such abundant shapes in nature, observable even on molecular levels. In fact, according to the results, any homogeneous closed surface with time invariable surface tension adopts CMC shape when it comes in equilibrium with environment.

\section*{Acknowledgments}

We thank Dr. Frank Julicher (MPIPKS) and Dr. Erwin Frey (LMU), 
for stimulating discussions. 
The paper in it's current form was initiated at Aspen Center for Physics, which is supported by National Science Foundation grant PHY-1607611 and was partially supported by a grant from the Simons Foundation. 

\section*{Appendix: Derivation of Equations of Motions for Closed, Two dimensional Surfaces}
Now we turn to the derivation of (\ref{41}-\ref{43}) without using any information from (\ref{38}-\ref{40}) (though derivation of  (\ref{41}-\ref{43}) from  (\ref{38}-\ref{40}) is strightforwad and trivial if one sets $V^0=0$ in (\ref{38}-\ref{40} equations \cite{2016arXiv160907765S}). To deduce the equations of motion we derive the simplest one from the set (\ref{41}) first. It is direct
consequence of generalization of conservation of mass law. Variation of the surface mass density must be so that $dm/dt=0$, where $m=\int_S\rho dS$ is surface mass with $\rho$ surface density. Since the surface is closed, at the boundary condition $v=n_iV^i=0$, a pass integral along any curve $\gamma$ across the surface must vanish ($n_i$ is a normal of the curve and lays in the tangent space). Using Gauss theorem, conservation of mass and integration formula (\ref{36}), we find
\begin{align} \label{48}
0=&\int_\gamma v\rho d\gamma=\int_\gamma n_iV^i\rho d\gamma=\int_S\nabla_i(\rho V^i)dS \nonumber \\
=&\int_S(\nabla_i(\rho V^i)-\rho CB_i^i+\rho CB_i^i)dS\nonumber \\
=&\int_S(\nabla_i(\rho V^i)-\rho CB^i_i)dS+\int_S\dot\nabla\rho dS-\frac{d}{dt}\int_S\rho dS \nonumber \\
=&\int_S(\dot\nabla\rho+\nabla_i(\rho V^i)-\rho CB_i^i)dS
\end{align}
Since last integral must be identical to zero for any integrand, one immediately finds first equation from the set (\ref{41}). To deduce second and third equations, we take a Lagrangian
\begin{equation}
L=\int_S\frac{\rho V^2}{2}dS+\int_\Omega(P^++\Pi)d\Omega
\end{equation}
and set minimum action principle requesting that $\delta L/\delta t=0$. Evaluation of space integral is simple and straightforward, using integration theorem for space integral where the convective and advective terms due to volume motion is properly taken into account (\ref{35}), we find
\begin{equation} \label{50}
\frac{\delta}{\delta t}\int_\Omega(P^++\Pi)d\Omega=\int_\Omega \partial_\alpha(P^++\Pi)V^\alpha d\Omega+\int_SC(P^++\Pi)dS
\end{equation}
Derivation for kinetic part is a bit tricky and challenging that is why we do it last. Straightforward, brute mathematical manipulations, using first equation from (\ref{41}), lead
\begin{align}
&\frac{\delta}{\delta t}\int_S\frac{\rho V^2}{2}dS=\int_S(\dot\nabla\frac{\rho V^2}{2}-CB_i^i\frac{\rho V^2}{2})dS \nonumber \\
&=\int_S(\dot\nabla\rho\frac{V^2}{2}+\rho\dot\nabla\frac{V^2}{2}-CB_i^i\frac{\rho V^2}{2})dS\nonumber \\
&=\int_S((\rho CB_i^i-\nabla_i(\rho V^i))\frac{V^2}{2}+\rho\dot\nabla\frac{V^2}{2}-CB_i^i\frac{\rho V^2}{2})dS \nonumber \\
&=\int_S(-\nabla_i(\rho V^i)\frac{V^2}{2}+\rho\dot\nabla\frac{V^2}{2})dS \nonumber \\
&=\int_S(-\nabla_i(\rho V^i\frac{V^2}{2})+\rho V^i\nabla_i\frac{V^2}{2}+\rho\dot\nabla\frac{V^2}{2})dS \nonumber \\
&=\int_S(-\nabla_i(\rho V^i\frac{V^2}{2})+\rho\vec{V}(V^i\nabla_i\vec{V}+\dot\nabla\vec{V}))dS
\end{align}
At the end point of variations the surface reaches stationary point and therefore by Gauss theorem (as we used it already in (\ref{48})), we find
\begin{equation}\label{52}
\int_S-\nabla_i(\rho V^i\frac{V^2}{2})dS=-\int_\gamma\rho V^in_i\frac{V^2}{2}d\gamma=0
\end{equation}
$\gamma$ is stationary contour of the surface and $n_i$ is the normal to the contour, therefore interface velocity for contour $v=n_iV^i=0$ and the integral (\ref{52}) vanishes, correspondingly
\begin{equation}\label{53}
\frac{\delta}{\delta t}\int_S\frac{\rho V^2}{2}dS=\int_S\rho\vec{V}(V^i\nabla_i\vec{V}+\dot\nabla\vec{V})dS
\end{equation}
To decompose dot product in the integral by normal and tangential components and, therefore, deduce final equations, we do following algebraic manipulations
\begin{widetext}
\begin{align}
\dot\nabla\vec{V}+V^i\nabla_i\vec{V}&=\dot\nabla\vec{V}+V^i\nabla_i\vec{V}+CV^iB_i^j\vec{S}_j-CV^iB_i^j\vec{S}_j=\dot\nabla\vec{V}+V^i\nabla_i\vec{V}+CV^iB_i^jX_j^\alpha\vec{X}_\alpha-CV^iB_i^j\vec{S}_j
\end{align}
\end{widetext}
Now using Weingarten’s formula $X_j^\alpha B_i^j=-\nabla_iN^\alpha$, metrinilic property of Euclidian space base vectors $\nabla_i\vec{X}_\alpha=0$, definition of surface normal $\vec{N}=N^\alpha\vec{X}_\alpha$ and taking into account definition of surface velocity $\vec{V}=C\vec{N}+V^i\vec{S}_i$ and its derivatives, we find 
\begin{widetext}
\begin{align} \label{55}
\dot\nabla\vec{V}+V^i\nabla_i\vec{V}+CV^iB_i^jX_j^\alpha\vec{X}_\alpha-CV^iB_i^j\vec{S}_j&=\dot\nabla\vec{V}+V^i\nabla_i\vec{V}-CV^i\vec{X}_\alpha\nabla_iN^\alpha-CV^iB_i^j\vec{S}_j \nonumber \\
&=\dot\nabla\vec{V}+V^i\nabla_i\vec{V}-CV^i\nabla_i(N^\alpha\vec{X}_\alpha)-CV^iB_i^j\vec{S}_j \nonumber \\
&=\dot\nabla\vec{V}+V^i\nabla_i\vec{V}-CV^i\nabla_i\vec{N}-CV^iB_i^j\vec{S}_j\nonumber \\
&=\dot\nabla\vec{V}+V^i\nabla_i(C\vec{N})+V^i\nabla_i(V^j\vec{S}_j)-CV^i\nabla_i\vec{N}-CV^iB_i^j\vec{S}_j\nonumber \\
&=\dot\nabla\vec{V}+V^i\vec{N}\nabla_iC+V^i\nabla_i(V^j\vec{S}_j)-CV^iB_i^j\vec{S}_j\nonumber \\
&=\dot\nabla(C\vec{N})+\dot\nabla(V^j\vec{S}_j)+V^i\vec{N}\nabla_iC+V^i\nabla_i(V^j\vec{S}_j)-CV^iB_i^j\vec{S}_j
\end{align}
\end{widetext}
Continuing algebraic manipulations using Thomas formula $\dot\nabla\vec{N}=-\nabla^iC\vec{S}_i$, the formula for surface derivative of interface velocity
$\vec{N}\nabla_iC=\dot\nabla\vec{S}_i$ and the definition of curvature tensor (\ref{23}) yield
\begin{widetext}
\begin{align} \label{56}
&\dot\nabla(C\vec{N})+\dot\nabla(V^j\vec{S}_j)+V^i\vec{N}\nabla_iC+V^i\nabla_i(V^j\vec{S}_j)-CV^iB_i^j\vec{S}_j \nonumber \\
&=\dot\nabla(C\vec{N})+C\nabla^jC\vec{S}_j+2V^i\vec{N}\nabla_iC+V^iV^jB_{ij}\vec{N}+\dot\nabla(V^j\vec{S}_j)\nonumber \\
&-V^i\vec{N}\nabla_iC+V^i\nabla_i(V^j\vec{S}_j)-V^iV^jB_{ij}\vec{N}-C\nabla^jC\vec{S}_j-CV^iB_i^j\vec{S}_j\nonumber \\
&=\dot\nabla(C\vec{N})-C\dot\nabla\vec{N}+2V^i\vec{N}\nabla_iC+V^iV^jB_{ij}\vec{N}+\dot\nabla(V^j\vec{S}_j)-V^j\dot\nabla\vec{S}_j \nonumber \\
&+V^i\nabla_i(V^j\vec{S}_j)-V^iV^j\nabla_i\vec{S}_j-C\nabla^jC\vec{S}_j-CV^iB^j_i\vec{S}_j\nonumber \\
&=(\dot\nabla C+2V^i\nabla_iC+V^iV^jB_{ij})\vec{N}+(\dot\nabla V^j+V^i\nabla_iV^j-C\nabla^jC-CV^iB_i^j)\vec{S}_j
\end{align}
\end{widetext}
Taking dot product of (\ref{56}) on $\vec{V}$ and combining it with (\ref{53}) last derivation finally reveals variation of kinetic energy, so that we finally find
\begin{widetext}
\begin{align}
 \label{57}
\frac{\delta}{\delta t}\int_S\frac{\rho V^2}{2}dS=\int_S(\rho C(\dot\nabla C+2V^i\nabla_iC+V^iV^jB_{ij})+\rho V_i(\dot\nabla V^i+V^j\nabla_jV^i-C\nabla^iC-CV^jB_j^i))dS
\end{align}
\end{widetext}
Combining (\ref{48}-\ref{50}) and (\ref{57}) together and taking into account that the pressure acts on the surface along the surface normal, we immediately find first (\ref{41}) and the last equation (\ref{43}) of the set. To clarify second equation (\ref{42}), we have
\begin{widetext}
\begin{align}
\int_S\rho C(\dot\nabla C+2V^i\nabla_iC+V^iV^jB_{ij})dS&=\int_\Omega-\partial_\alpha(P^++\Pi)V^\alpha d\Omega-\int_SC(P^++\Pi)dS \nonumber \\
\int_SC(\rho(\dot\nabla C+2V^i\nabla_iC+V^iV^jB_{ij})+P^++\Pi)dS&=\int_\Omega-\partial_\alpha(P^++\Pi)V^\alpha d\Omega \label{58}
\end{align}
\end{widetext}
After applying Gauss theorem to the second equation (\ref{58}), the surface integral is converted to space integral so that we finally find
\begin{widetext}
\begin{align}
\partial_\alpha(\rho V^\alpha(\dot\nabla C+2V^i\nabla_iC+V^iV^jB_{ij})+(P^++\Pi)V^\alpha)=-\partial_\alpha(P^++\Pi)V^\alpha
\end{align}
\end{widetext}
and, therefore, all three equations (\ref{41}-\ref{43}) are rigorously clarified.

\bibliography{sample}

\end{document}